\documentclass[10pt,conference,letterpaper]{IEEEtran}
\usepackage[USenglish]{babel}
\usepackage{times,amsmath}
\usepackage[pdftex]{graphicx}
   \graphicspath{{./pdf/}{./png/}{./jpeg/}{./eps/}{./png_old/}}
   \DeclareGraphicsExtensions{.pdf,.png,.jpeg,.eps}
\usepackage{cite}
\usepackage{url}
\usepackage{fmtcount}
\usepackage{color}
\usepackage[tight,footnotesize]{subfigure}
\usepackage{multicol}
\usepackage{amssymb}
\usepackage{bm}
\usepackage{wasysym}
\usepackage{dblfloatfix} % to enable figures at the bottom of page
\usepackage{calc}
\usepackage{fix-cm}
\usepackage[activate={true,nocompatibility},final,tracking=true,factor=500,stretch=40,shrink=40]{microtype}
\usepackage{stmaryrd}
\usepackage{textcomp}
\usepackage{mathtools}
\usepackage[libertine]{newtxmath}
\renewcommand{\d}{{\mathrm d}}

\newcommand{\half}{{\frac{1}{2}}}
\newcommand{\Ibl}{{\llbracket}}
\newcommand{\Ibr}{{\rrbracket}}
\renewcommand{\mod}{{\rm mod}}

\DeclareMathOperator\erfc{erfc}
\newcommand{\appropto}{\mathrel{\vcenter{
  \offinterlineskip\halign{\hfil$##$\cr
    \propto\cr\noalign{\kern2pt}\sim\cr\noalign{\kern-2pt}}}}}
\definecolor{DarkRed}{rgb}{0.5,0,0}
\definecolor{DarkGreen}{rgb}{0,0.5,0}
\definecolor{DarkerGreen}{rgb}{0,0.3333,0}
\definecolor{DarkBlue}{rgb}{0,0,0.75}
\definecolor{RoyalBlue}{rgb}{0,0.1373,0.4000}
\definecolor{NavyBlue}{rgb}{0,0,0.5020}
\definecolor{CobaltBlue}{rgb}{0,0.2784,0.6706}
\definecolor{lightlightgray}{rgb}{0.96875,0.96875,0.96875}
\definecolor{cyan}{rgb}{0,1,1}
\newcommand{\beginlabel}[2]{%
\begin{#1}\label{#2}}
\begin{document}
\pagestyle{plain}
%\pagestyle{headings}
%--------------------------------------------------------------------------
% TITLE
\title{Utilizing Pulse Pileup Effect in Development of Robust Low-SNR Covert Communication Links}
\author{\IEEEauthorblockN{Alexei V. Nikitin}
\IEEEauthorblockA{%\\
Nonlinear LLC\\
Wamego, Kansas, USA\\
E-mail: avn@nonlinearcorp.com}
\and
\IEEEauthorblockN{Ruslan L. Davidchack}
\IEEEauthorblockA{Dept. of Mathematics, U. of Leicester\\
Leicester, UK\\
E-mail: rld8@leicester.ac.uk}}
%--------------------------------------------------------------------------
\maketitle
%\thispagestyle{plain}
%--------------------------------------------------------------------------
% ABSTRACT
\begin{abstract}
In contrast to other spread-spectrum techniques, wideband pulse trains with relatively low pulse arrival rates may be considered unsuitable for covert communications. The high crest factor of such trains can be extremely burdensome for the transmitter hardware, and it makes the pulse trains easily detectable even at very low signal-to-noise ratios. In addition, it may appear that sharing the wideband channel by multiple users would require explicit allocation of the pulse arrival times for each sub-channel, which would be impractical in most cases. On the other hand, messaging by wideband pulse trains has many appealing features. Among those are the ease of synchronous as well as asynchronous pulse detection, and on-the-fly channel reconfigurability (e.g. changing the spreading factor). Favorably, the crest factor of a pulse train, as well as its apparent temporal and amplitude structure, can be easily, and reversibly, controlled by simple linear filtering. For example, a transmitted pulse train can be made statistically indistinguishable from the Gaussian component of the channel noise (e.g. the thermal noise) observed in the same spectral band, while the received signal will be the designed high-crest-factor wideband pulse train. In this paper, we utilize the so-called pulse pileup effect to perform such reversible control of the pulse train structure, enabling a wider use of this approach for synthesis of robust low-SNR covert communication links. We place a particular focus on the synchronous pulse detection in the receiver, that provides a better utilization of the channel spectrum.
\end{abstract}
%--------------------------------------------------------------------------
%--------------------------------------------------------------------------
\maketitle
%--------------------------------------------------------------------------

%--------------------------------------------------------------------------
% FIGURE -- Pulse Trains for Low-SNR Communications
%%% placed here to appear at top of next page
% up to 17.8 cm full width
%--------------------------------------------------------------------------
\begin{figure*}[!t]
\centering{\includegraphics[width=17.4cm]{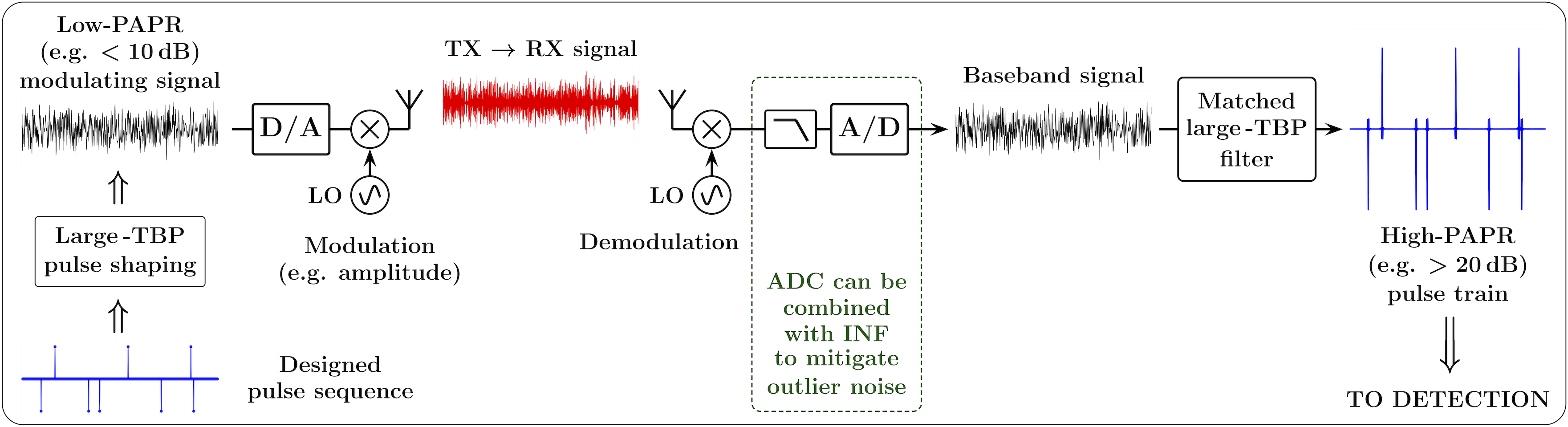}}
\caption{Using pulse trains for low-SNR communications: Large-TBP pulse shaping (i)~``hides" pulse train, obscuring its temporal and amplitude structure, and (ii)~reduces its PAPR, making signal suitable for transmission. In receiver, pulse train is restored by matched large-TBP filtering. High PAPR of restored pulse train enables low-SNR messaging. To make link more robust to outlier interference and to increase apparent SNR, analog-to-digital conversion in receiver can be combined with intermittently nonlinear filtering.
\label{fig:low SNR}}
\end{figure*}
%--------------------------------------------------------------------------

%--------------------------------------------------------------------------
% INTRODUCTION
\section{Introduction} \label{sec:introduction}
%--------------------------------------------------------------------------
The additive white Gaussian noise (AWGN) capacity~$C$ of a channel operating in the power-limited regime (i.e. when the received signal-to-noise ratio (SNR) is small, ${\rm SNR} \ll 0\,$dB) can be expressed as $C\approx \bar{P}/(N_0\,\ln 2)$, where~$\bar{P}$ is the average received power and~$N_0$ is the power spectral density (PSD) of the noise. This capacity is linear in power and insensitive to bandwidth and, therefore, by spreading the average transmitted power of the information-carrying signal over a large frequency band, the average PSD of the signal could be made much smaller than the PSD of the noise. This would ``hide" the signal in the channel noise, making the transmission covert and insensitive to narrowband interference.

One of the common ways to achieve such ``spreading" is frequency-hopping spread spectrum (FHSS)~\cite{Simon94spreadspectrum}. This technique is widely used, for example, in legacy military equipment for low-probability-of-intercept (LPI) communications. However, using frequency hopping for covert communications is nearly obsolete today, since modern wideband software-defined radio (SDR) receivers can capture all of the hops and put them back together (J.~E.~Gilley, personal communication, Feb. 9, 2020).

Another common and widely used spread-spectrum modulation technique is direct-sequence spread spectrum (DSSS)~\cite{Torrieri18principles}. In DSSS, the narrow-band information-carrying signal of a given power is modulated by a wider-band, unit-power pseudorandom signal known as a spreading sequence. This results in a signal with the same total power but a larger bandwidth, and thus a smaller PSD. After demodulation (``de-spreading") in the receiver, the original information-carrying signal is restored. However, such demodulation requires a precise synchronization, which is perhaps the most difficult and expensive aspect of a DSSS receiver design. Also, while de-spreading cannot be performed without the knowledge of the spreading sequence by the receiver, the spreading code by itself may not be usable to secure the channel. For example, linear spreading codes are easily decipherable once a short sequential set of chips from the sequence is known. To improve security, it would be desirable to perform a ``code hopping" in a manner akin to the frequency hopping. However, synchronization can be an extremely slow process for pseudorandom sequences, especially for large spreading waveforms (long codes), and thus such DSSS code hopping may be difficult to realize in practice.

In this paper, we explore an alternative spread-spectrum approach that, among other appealing features, significantly simplifies and speeds up synchronization and enables on-the-fly link reconfigurability that combines the benefits of both DSSS and FHSS. A simplified explanation of this approach can be given as follows.

In the power-limited regime, we would normally use binary coding and modulation (e.g. binary phase-shift keying (BPSK) or quadrature phase-shift keying (QPSK)) for the narrow-band information-carrying signal, and this signal will be significantly oversampled to enable wideband spreading. Thus an idealized narrow-band information-carrying signal that is to be ``spread" can be viewed as a discrete-level signal that is a linear combination of analog Heaviside unit step functions~\cite{Bracewell2000Fourier} delayed by multiples of the bit duration. Such a signal would have a limited bandwidth and a finite power. Since the derivative of the Heaviside unit step function is the Dirac $\delta$-function~\cite{Dirac58principles}, the derivative of this idealized signal will be a ``pulse train" that is a linear combination of Dirac $\delta$-functions. This pulse train will contain all the information encoded in the discrete-level signal, and it will have infinitely wide bandwidth and infinitely large power. Both the bandwidth and the power can then be reduced to the desired levels by filtering the pulse train with a lowpass filter. If the time-bandwidth product (TBP) of the filter is sufficiently small so that the pulses in the filtered pulse train do not overlap, these pulses will still contain all the intended information.

On the one hand, converting a narrow-band signal into a wideband pulse train has an apparent appeal of no need for ``de-spreading": One can simply obtain samples at the peaks of the pulses to obtain all the information encoded in the signal. On the other hand, at first glance such a pulse train is not suitable for practical communication systems, and especially for covert communications. Indeed, let us consider a pulse train with a given average pulse rate and power. The average PSD of this train can be made arbitrary small, since it is inversely proportional to the bandwidth. However, the peak-to-average power ratio (PAPR) of such a train would be proportional to the bandwidth, making the wideband signal extremely impulsive (super-Gaussian). First, such high crest factor of the pulse train puts a serious burden on the transmitter hardware, potentially making this burden prohibitive (e.g. for ${\rm PAPR}>30\,$dB). Secondly, the high-PAPR structure of a pulse train makes it easily detectable by simple thresholding in the time domain, seemingly making it unsuitable for covert communications. Thirdly, it may appear that sharing the wideband channel by multiple users would require explicit allocation of the pulse arrival times for each sub-channel, which would be impractical in most cases.

Favorably, the temporal and amplitude structure of a pulse train is modifiable by linear filtering, and such filtering can convert a high-PAPR train into a low-PAPR signal, and {\em vice versa\/}. Therefore, such PAPR-modifying filtering enables us to use pulse trains for low-SNR covert communications. As illustrated in Fig.~\ref{fig:low SNR}, large-TBP pulse shaping in the transmitter can ``hide" the pulse train, obscuring its temporal and amplitude structure. It also reduces the PAPR of the signal, making it suitable for transmission. In the receiver, the distinct structure of the pulse train is restored by matched large-TBP filtering, and the high PAPR of the restored pulse train enables low-SNR messaging. To make such a link more robust to outlier interference and to increase the apparent SNR, analog-to-digital conversion in the receiver can be combined with intermittently nonlinear filtering (INF)~\cite{Nikitin19hidden, Nikitin19complementary}.

%--------------------------------------------------------------------------
% FIGURE
%--------------------------------------------------------------------------
\begin{figure}[!b]
\centering{\includegraphics[width=7.8cm]{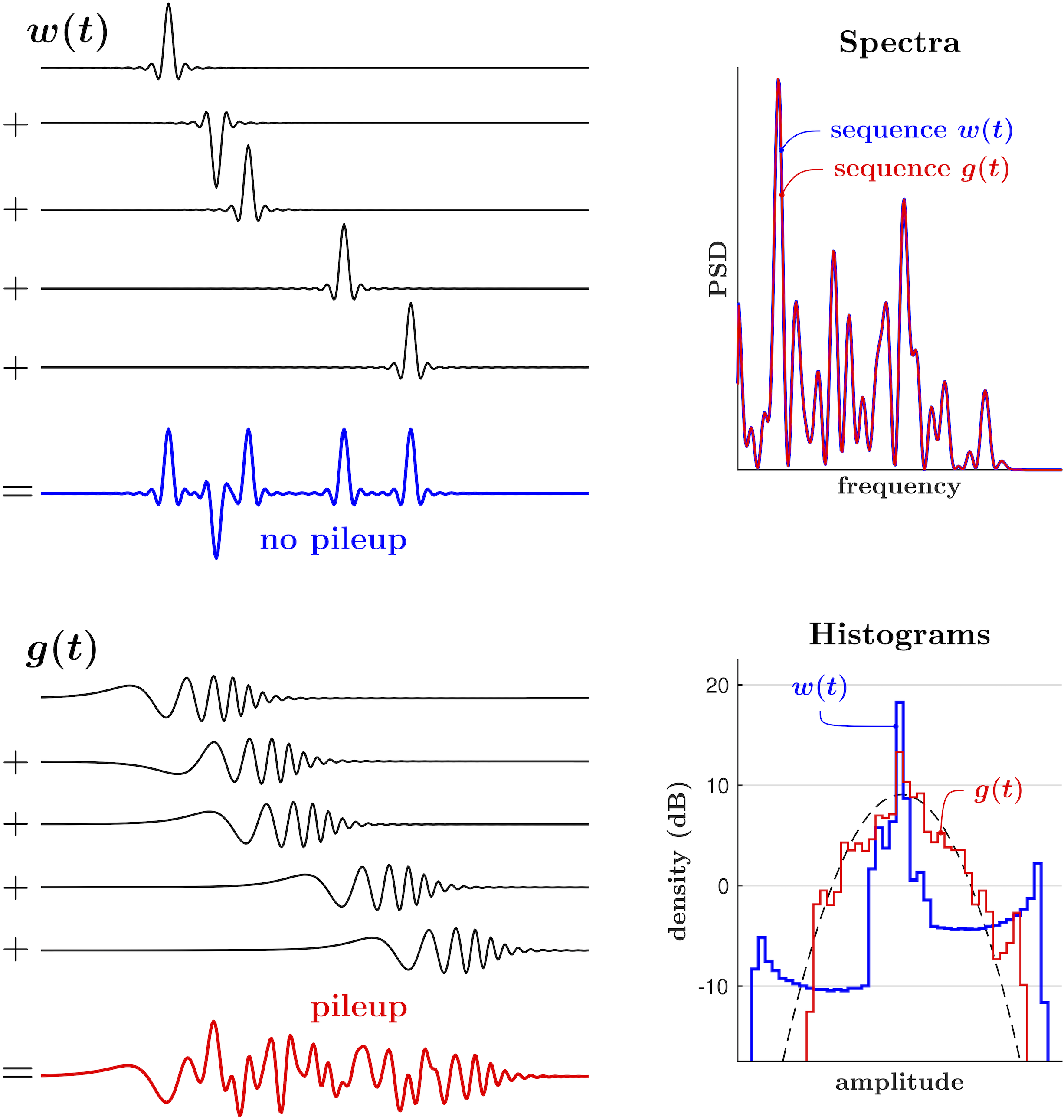}}
\caption{Illustration of pileup effect: When ``width" of pulses becomes greater than distance between them, pulses begin to overlap and interfere with each other. For pulses with same spectral content, PSDs of pulse sequences are identical, yet their temporal and amplitude structures are substantially different.
\label{fig:PPileup}}
\end{figure}
%--------------------------------------------------------------------------

The focus of the rest of the paper is on the key components of such a link, and on the synchronous pulse detection in particular. Some of the additional aspects of this approach, such as asynchronous detection, multi-layer and multi-user configurations, and applications for physical-layer steganography and ``friendly jamming" are outlined in~\cite{Nikitin20steganography}.

%--------------------------------------------------------------------------
% PAPR Control Utility of Pileup Effect
\section{PAPR Control Utility of Pileup Effect}
%--------------------------------------------------------------------------
A pulse train~$p(t)$ is simply a sum of pulses with the same shape (impulse response)~$w(t)$, same or different amplitudes~$a_k$, and distinct arrival times~$t_k$: $p(t) = \sum_k a_k w(t-t_k)$. When the width of the pulses in a train becomes greater than the distance between them, the pulses begin to overlap and interfere with each other. This is illustrated in Fig.~\ref{fig:PPileup}: For the same arrival times, the pulses in the sequence consisting of the narrow pulses~$w(t)$ remain separate, while the wider (more ``spread out") pulses~$g(t)$ are ``piling up on top of each other." In this example, $w(t)$ and $g(t)$ have the same spectral content, and thus the PSDs of the pulse sequences are identical. However, the ``pileup effect" causes the temporal and amplitude structures of these sequences to be substantially different. For a random pulse train, when the ratio of the bandwidth and the pulse arrival rate becomes significantly smaller than the TBP of a pulse, the pileup effect causes the resulting signal to become effectively Gaussian~\cite[e.g.]{Nikitin98ppileup}, making it impossible to distinguish between the individual pulses.

Indeed, let~$\hat{p}(t)$ be an ``ideal" pulse train: $\hat{p}(t) = \sum_k a_k\delta(t-t_k)$, where $\delta(x)$ is the Dirac $\delta$-function~\cite{Dirac58principles}. The {\itshape moving average\/} of this ideal train in a boxcar window of width~$2T$ can be represented by the convolution integral
%--------------------------------------------------------------------------
\begin{equation} \label{eq:moving average}
  \overline{p}(t) = \int_{-\infty}^\infty \!\!\d{s}\, \frac{\theta(t\!+\!T)-\theta(t\!-\!T)}{2T}\, \hat{p}(t\!-\!s)\,,
\end{equation}
%--------------------------------------------------------------------------
where $\theta(x)$ is the Heaviside unit step function~\cite{Bracewell2000Fourier}. At any given time~$t_i$, the value of $\overline{p}(t_i)$ is proportional to the sum of~$a_k$ for the pulses that occur within the interval~$[t_i\!-\!T,t_i\!+\!T]$. Then, if the amplitudes $a_k$ and/or the interarrival times~$t_{k+1}-t_k$ are independent and identically distributed (i.i.d.) random variables with finite mean and variance, it follows from the Central Limit Theorem~\cite[e.g.]{Aleksandrov56mathematics} that the distribution of $\overline{p}(t_i)$ approaches Gaussian for a sufficiently large interval~$[-T,T]$.

If we replace the boxcar weighting function in~(\ref{eq:moving average}) with an arbitrary moving window~$w(t)$, then~(\ref{eq:moving average}) becomes a {\itshape weighted\/} moving average
%--------------------------------------------------------------------------
\begin{equation} \label{eq:weighted moving average}
  p(t) = \int_{-\infty}^\infty \!\!\d{s}\, w(t)\, \hat{p}(t\!-\!s) = (\hat{p}\!\ast\!w)(t) = \sum_k a_k w(t-t_k)\,,
\end{equation}
%--------------------------------------------------------------------------
%where the asterisk denotes convolution,
%
which is a ``real" pulse train with the impulse response~$w(t)$.
%\footnote{In~(\ref{eq:weighted moving average}) and throughout the paper the asterisk denotes convolution.}
If~$w(t)$ is normalized so that~$\int_{-\infty}^\infty \d{s}\, w(s) = 1$, $w(t)$ is an {\itshape averaging\/} (i.e. lowpass) filter. Then, if~$w(t)$ has both the bandwidth and the TBP similar to that of the boxcar pulse of width~$2T$, the distribution of $p(t_i)$ would be similar to that of~$\overline{p}(t_i)$ (e.g. Gaussian for a sufficiently large~$T$).

%--------------------------------------------------------------------------
% PAPR control by large-TBP pulse shaping
\subsection{PAPR Control by Large-TBP Pulse Shaping} \label{subsec:PAPR control}
%--------------------------------------------------------------------------
There are various ways to define the ``time duration" and the ``bandwidth" of a pulse. This can lead to a significant ambiguity in the definitions of the TBPs, especially for filters with complicated temporal structures and/or frequency responses. However, in the context of a PAPR control function of the pileup effect, our main concern is the change in the TBP that occurs only due to the change in the temporal structure of a filter, without the respective change in its spectral content. For a single pulse~$w(t)$, its PAPR can be expressed as
%--------------------------------------------------------------------------
\begin{equation} \label{eq:PAPR}
  {\rm PAPR}_w = \frac{\max\left(w^2(t)\right)}{\frac{1}{T_2-T_1} \int_{T_1}^{T_2} \d{t}\, w^2(t)}\,,
\end{equation}
%--------------------------------------------------------------------------
where the interval~$[T_1,T_2]$ includes the effective time support of~$w(t)$. Then for filters with the same spectral content and the impulse responses~$w(t)$ and~$g(t)$, the ratio of their TBPs can be expressed as the reciprocal of the ratio of their PAPRs,
%--------------------------------------------------------------------------
\begin{equation} \label{eq:TBP}
  \frac{{\rm TBP}_g}{{\rm TBP}_w} = \frac{\max\left(w^2(t)\right)}{\max\left(g^2(t)\right)} = \frac{{\rm PAPR}_w}{{\rm PAPR}_g}\,,
\end{equation}
%--------------------------------------------------------------------------
where the PAPRs are calculated over a sufficiently long time interval that includes the effective time support of both filters. Note that from~(\ref{eq:TBP}) it follows that, among all possible pulses with the same spectral content, the one with the smallest TBP will contain a dominating large-magnitude peak. Hence any reasonable definition of a finite TBP for a particular filter with a given frequency response allows us to obtain comparable numerical values for the TBPs of all other filters with the same frequency response, regardless of their temporal structures.

%--------------------------------------------------------------------------
% FIGURE -- allpass2chirp
%--------------------------------------------------------------------------
\begin{figure}[!b]
\centering{\includegraphics[width=8.6cm]{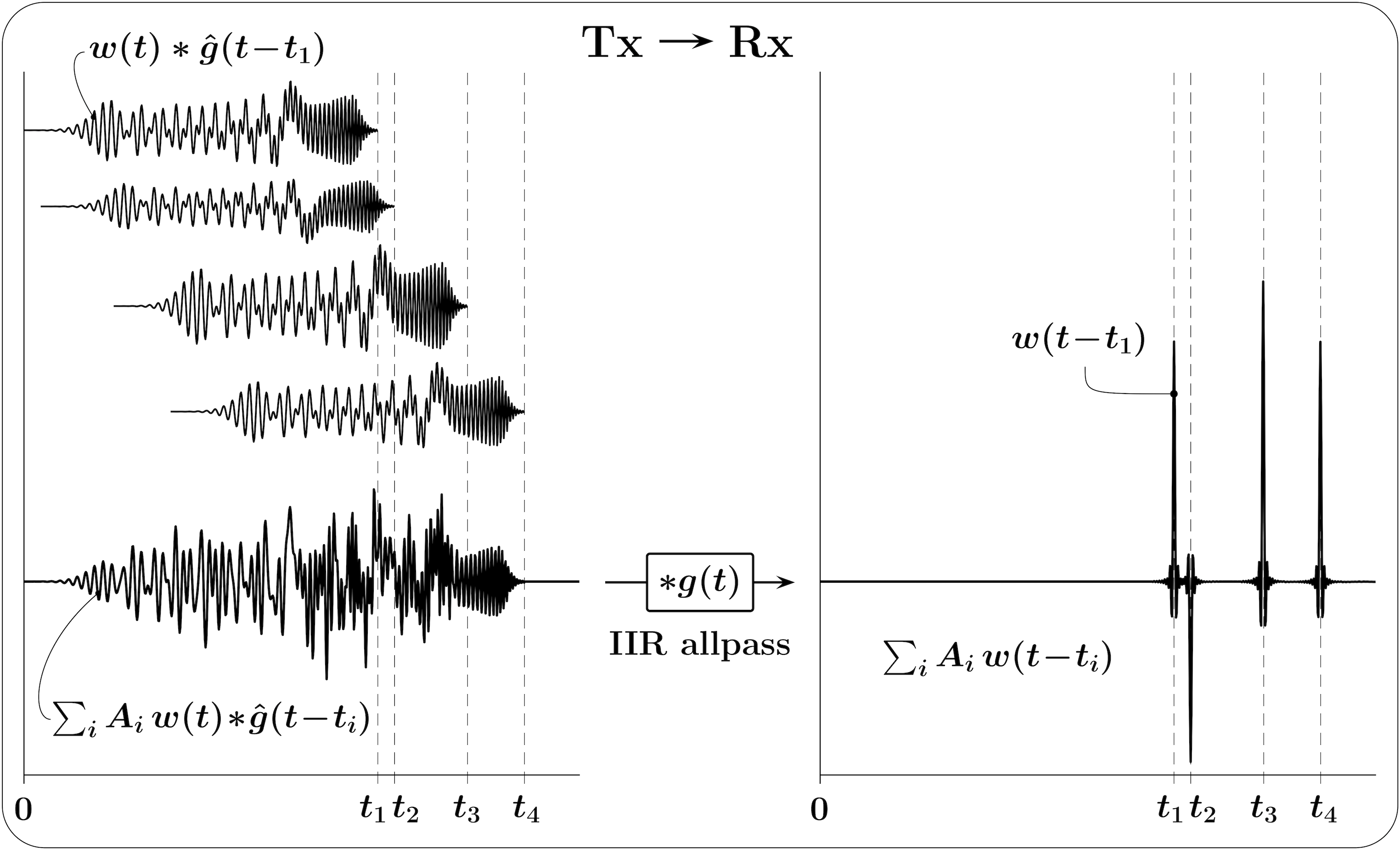}}
\caption{Transmitter waveform is constructed as sum of scaled and time-shifted large-TBP pulses. In receiver, IIR allpass filter recovers small-TBP pulse train.
\label{fig:allpass2chirp}}
\end{figure}
%--------------------------------------------------------------------------
%--------------------------------------------------------------------------
% FIGURE -- ppileup
%%% placed here to appear at top of next page
% up to 17.8 cm full width
%--------------------------------------------------------------------------
\begin{figure*}[!t]
\centering{\includegraphics[width=14cm]{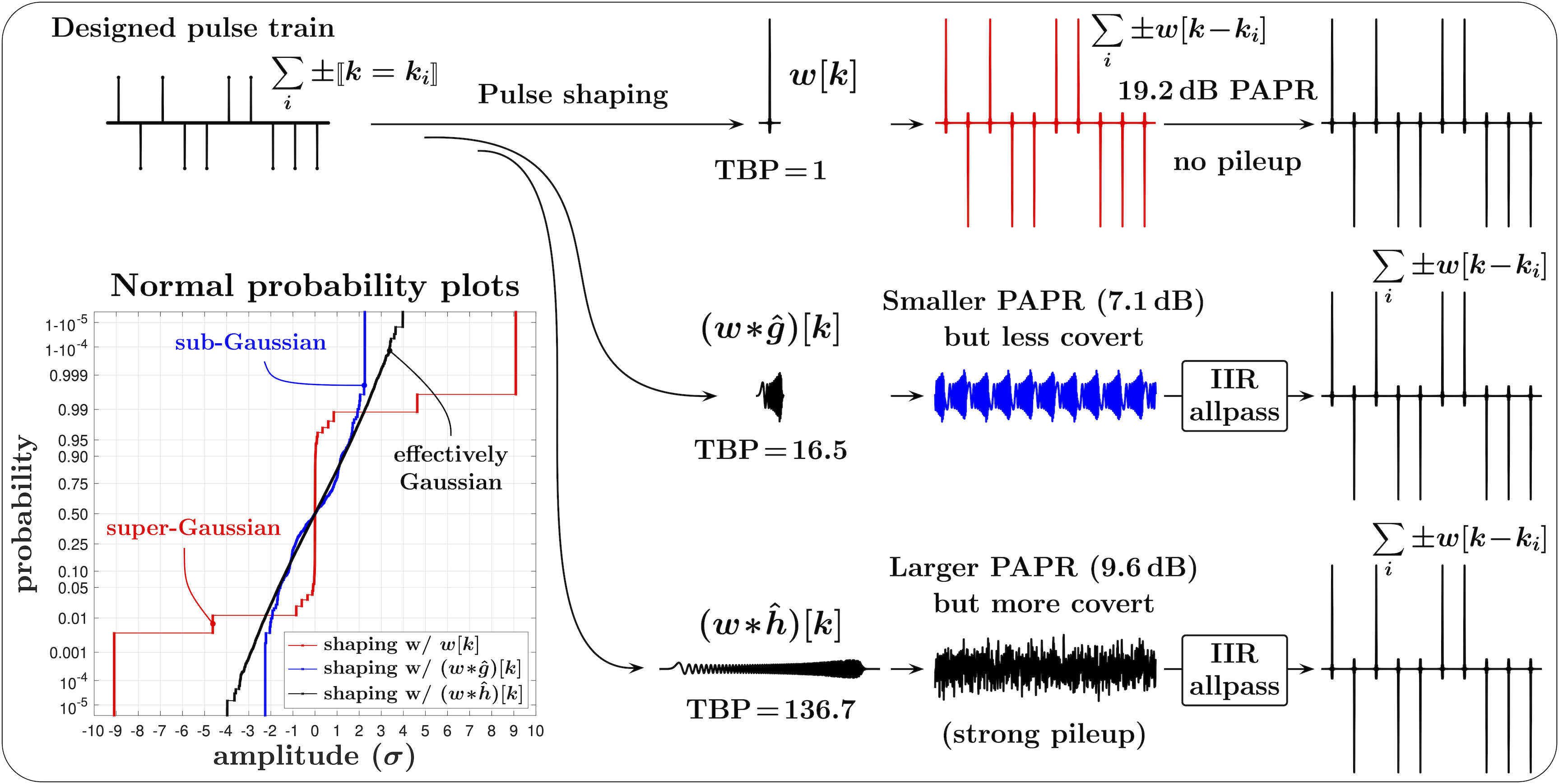}}
\caption{Using large-TBP filtering and pileup effect for obfuscation of temporal and amplitude structure of transmitted signal. In transmitter, pulse shaping with large-TBP filter reduces crest factor of pulse train, making it appear as sub-Gaussian or effectively Gaussian. In receiver, signal's distinct temporal and amplitude structure is restored.
\label{fig:ppileup}}
\end{figure*}
%--------------------------------------------------------------------------

There are multiple ways to construct pulses with identical frequency responses yet significantly different TBPs. For example, given a ``seed" small-TBP pulse with finite (FIR) or infinite (IIR) impulse response~$w(t)$, a large-TBP pulse with the same spectral content can be ``grown" from~$w(t)$ by applying a sequence of IIR allpass filters that leave the PSD of the seed pulse unmodified~\cite[e.g.]{Regalia88digital}. Then an FIR filter for pulse shaping in the transmitter can be obtained by (i)~``spreading"~$w(t)$ with an IIR allpass filter, (ii)~truncating the pulse when it sufficiently decays to zero, and (iii)~time-inverting the resulting waveform. Then applying the same IIR allpass filter in the receiver to this waveform will produce the matched filter $w(-t)$ to the original seed pulse.

In the illustration of Fig.~\ref{fig:allpass2chirp}, the transmitter waveform is composed as a ``piled-up" sum of thus constructed large-TBP pulses, scaled and time-shifted. In the receiver, an IIR allpass filter recovers the underlying high-PAPR pulse train. The seed~$w(t)$ used in this illustration is an FIR root-raised-cosine (RRC) pulse symmetrical around $t\!=\!0$, and thus $(w\!\ast\! w)(t)$ is a raised-cosine (RC) pulse. RC pulses are perhaps not the best choice for shaping the pulse trains for communications, since their TBP is only about unity, and pulse shaping with Gaussian or Bessel filters (with ${\rm TBP}\approx 2\ln(2)/\pi \approx 0.44$) may provide a better alternative. In the subsequent simulations and numerical examples, however, we use FIR RC pulses with roll-off factor~$\beta\!=\!1/2$ for convenience of their well-defined bandwidth and numerical values associated with their symbol-rate.  

Fig.~\ref{fig:ppileup} further illustrates how the pileup effect can be used to obscure (e.g. to mimic as Gaussian or sub-Gaussian) a large-PAPR (super-Gaussian) transmitted signal, while fully recovering its distinct temporal and amplitude structure in the receiver. In this example, pulse shaping with a large-TBP filter in the transmitter ``hides" the original structure of the pulse train, and the pulses with larger TBPs perform this more effectively. This can be seen in Fig.~\ref{fig:ppileup} from both the time-domain traces and the normal probability plots shown in the lower left corner. For a sufficiently large TBP, the distribution of the filtered pulse train with random pulse polarities becomes effectively Gaussian, making it impossible to distinguish between the individual pulses.

%--------------------------------------------------------------------------
% Pulse Trains for Low-SNR Communications
\section{Pulse Trains for Low-SNR Communications} \label{sec:pulse trains}
%--------------------------------------------------------------------------
Having demonstrated how the information-carrying pulse train can be made transmittable and covert, we shall discuss how the message can be best recovered in the receiver.

%--------------------------------------------------------------------------
% Synchronous Pulse Detection
\subsection{Synchronous Pulse Detection} \label{subsec:synchronous}
%--------------------------------------------------------------------------
Let us consider a pulse train consisting of pulses with the bandwidth~$\Delta{B}$ and a small TBP, so that a single large-magnitude peak in a pulse dominates, and assume that the arrival rate~${\mathcal{R}}$ of the pulses is sufficiently small so that pileup is negligible (e.g. ${\mathcal{R}}\ll{\mathcal{R}}_0\!=\!\half\Delta{B}/{\rm TBP}$). When the arrival time of a pulse with the peak magnitude~$|A|$ is known, the probability of correctly detecting the polarity of this pulse in the presence of additive white Gaussian noise (AWGN) with zero mean and $\sigma_{\rm n}^2$ variance can be expressed, using the complementary error function, as ${\half\erfc\left(\frac{-|A|}{\sigma_{\rm n}\sqrt{2}}\right)}$. Then the pulses with the magnitude ${|A|>\sigma_{\rm n}\sqrt{2}\erfc^{-1}(2\varepsilon)}$ will have a pulse identification error rate smaller than~$\varepsilon$. For example, $\varepsilon\lesssim 1.3\!\times\! 10^{-3}$ for $|A|\gtrsim 3\sigma_{\rm n}$, and $\varepsilon\lesssim 3.2\!\times\! 10^{-5}$ for $|A|\gtrsim 4\sigma_{\rm n}$.

%--------------------------------------------------------------------------
% FIGURE -- PAPR of RC
%--------------------------------------------------------------------------
\begin{figure}[!b]
\centering{\includegraphics[width=8.6cm]{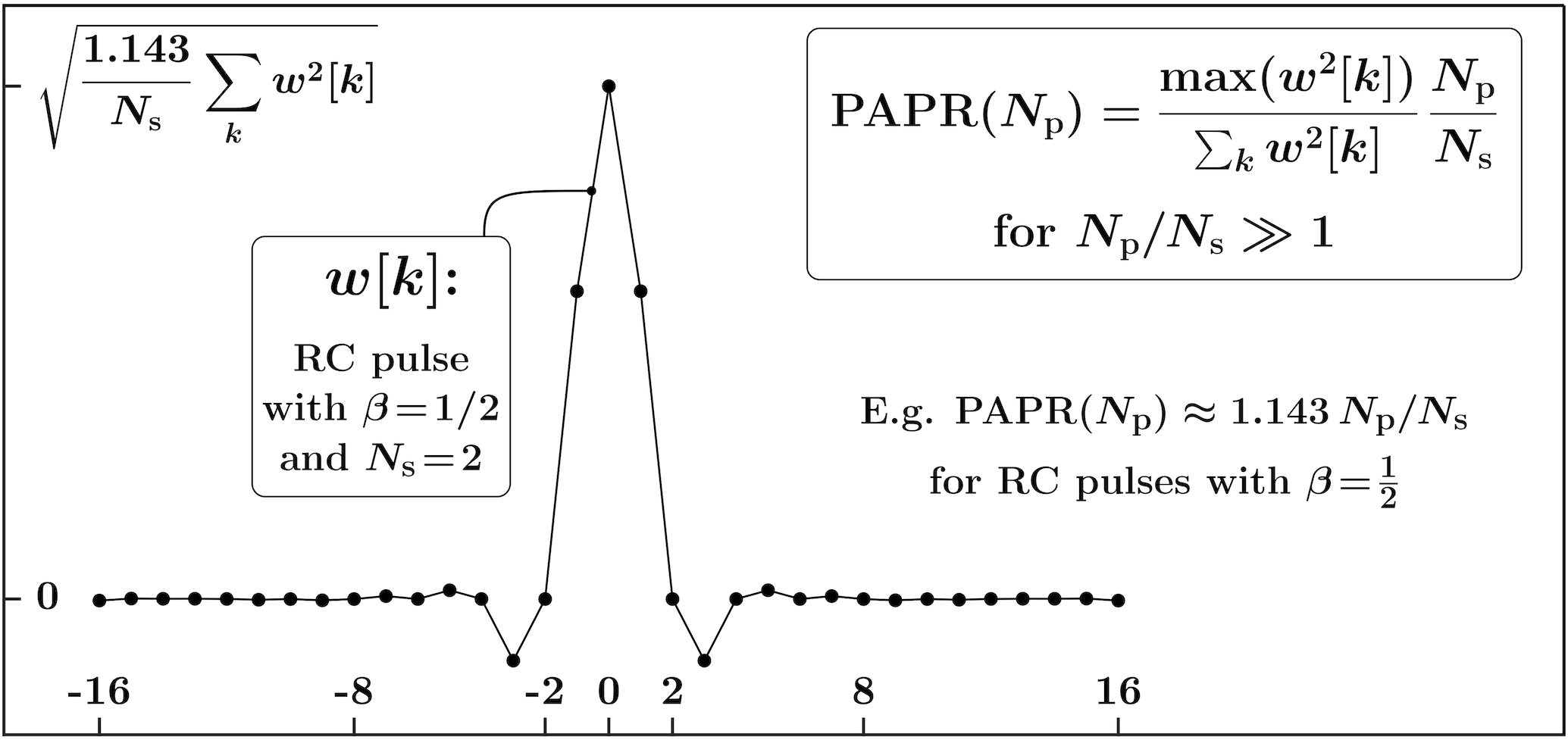}}
\caption{${{\rm PAPR}(N_{\rm p}) \approx 1.143\, N_{\rm p}/N_{\rm s}}$ for~$N_{\rm p}/N_{\rm s}\gg 1$ for RC pulses with $\beta\!=\!1/2$.
\label{fig:RC PAPR}}
\end{figure}
%--------------------------------------------------------------------------

The pulse rate in a digitally sampled train with regular (periodic) arrival times is $\mathcal{R}=F_{\rm s}/N_{\rm p}$, where $F_{\rm s}$ is the sampling frequency and $N_{\rm p}$ is the number of samples between two adjacent pulses in the train. For~${\mathcal{R}}$ that is sufficiently smaller than ${\mathcal{R}_0}$, the PAPR of a train of equal-magnitude pulses with regular arrival times is an {\em increasing\/} function of the number of samples between two adjacent pulses~${N_{\rm p}}$, and is proportional to~${N_{\rm p}}$:
%--------------------------------------------------------------------------
\begin{equation} \label{eq:PAPR vs rate}
  {\rm PAPR} = {\rm PAPR}(N_{\rm p}) \propto {N_{\rm p}} \quad \mbox{for large} \quad N_{\rm p}\,.
\end{equation}
%--------------------------------------------------------------------------
For example, for raised-cosine (RC) pulses ${\mathcal{R}}_0\approx (4T_{\rm s})^{-1}$, where $T_{\rm s}$ is the symbol-period, and a ``large $N_{\rm p}$" would mean ${N_{\rm p}\gg T_{\rm s}F_{\rm s} = N_{\rm s}}$, where $N_{\rm s}$ is the number of samples per symbol-period. As illustrated in Fig.~\ref{fig:RC PAPR}, ${{\rm PAPR}(N_{\rm p}) \approx 1.143\, N_{\rm p}/N_{\rm s}}$ for~$N_{\rm p}/N_{\rm s}\gg 1$ for RC pulses with roll-off factor~$\beta\!=\!1/2$.

From the lower limit on the magnitude of a pulse for a given uncoded bit error rate (BER),
%--------------------------------------------------------------------------
\begin{equation} \label{eq:A synch}
  |A| = \sigma_{\rm n} \sqrt{{\rm SNR}\!\times\! {\rm PAPR}} > \sigma_{\rm n}\sqrt{2}\erfc^{-1}(2\times{\rm BER})\,,
\end{equation}
%--------------------------------------------------------------------------
we can then obtain the lower limit on the SNR for a given pulse rate:
%--------------------------------------------------------------------------
\begin{equation} \label{eq:SNR vs Np}
  {\rm SNR}(N_{\rm p};{\rm BER}) > \frac{2\left[ \erfc^{-1}(2\times{\rm BER}) \right]^2}{{\rm PAPR}(N_{\rm p})} \propto {N_{\rm p}^{-1}},
\end{equation}
%--------------------------------------------------------------------------
or
%--------------------------------------------------------------------------
\begin{equation} \label{eq:SNR vs Np RC}
  {\rm SNR}(N_{\rm p};{\rm BER}) \gtrsim 1.75\left[ \erfc^{-1}(2\times{\rm BER}) \right]^2 \frac{N_{\rm s}}{N_{\rm p}}
\end{equation}
%--------------------------------------------------------------------------
for~$N_{\rm s}/N_{\rm p}\ll 1$ and RC pulses with~$\beta\!=\!1/2$. For example, ${\rm SNR}(N_{\rm p};10^{-3}) \gtrsim 9.6/{\rm PAPR}(N_{\rm p}) \approx 8.4\, N_{\rm s}/N_{\rm p}$, and ${\rm SNR}(N_{\rm p};10^{-5}) \gtrsim 18.2/{\rm PAPR}(N_{\rm p})  \approx 15.9\, N_{\rm s}/N_{\rm p}$.

%--------------------------------------------------------------------------
% FIGURE -- SNR limits
%%% placed here to appear at bottom of this column 
%--------------------------------------------------------------------------
\begin{figure}[!b]
\centering{\includegraphics[width=8.6cm]{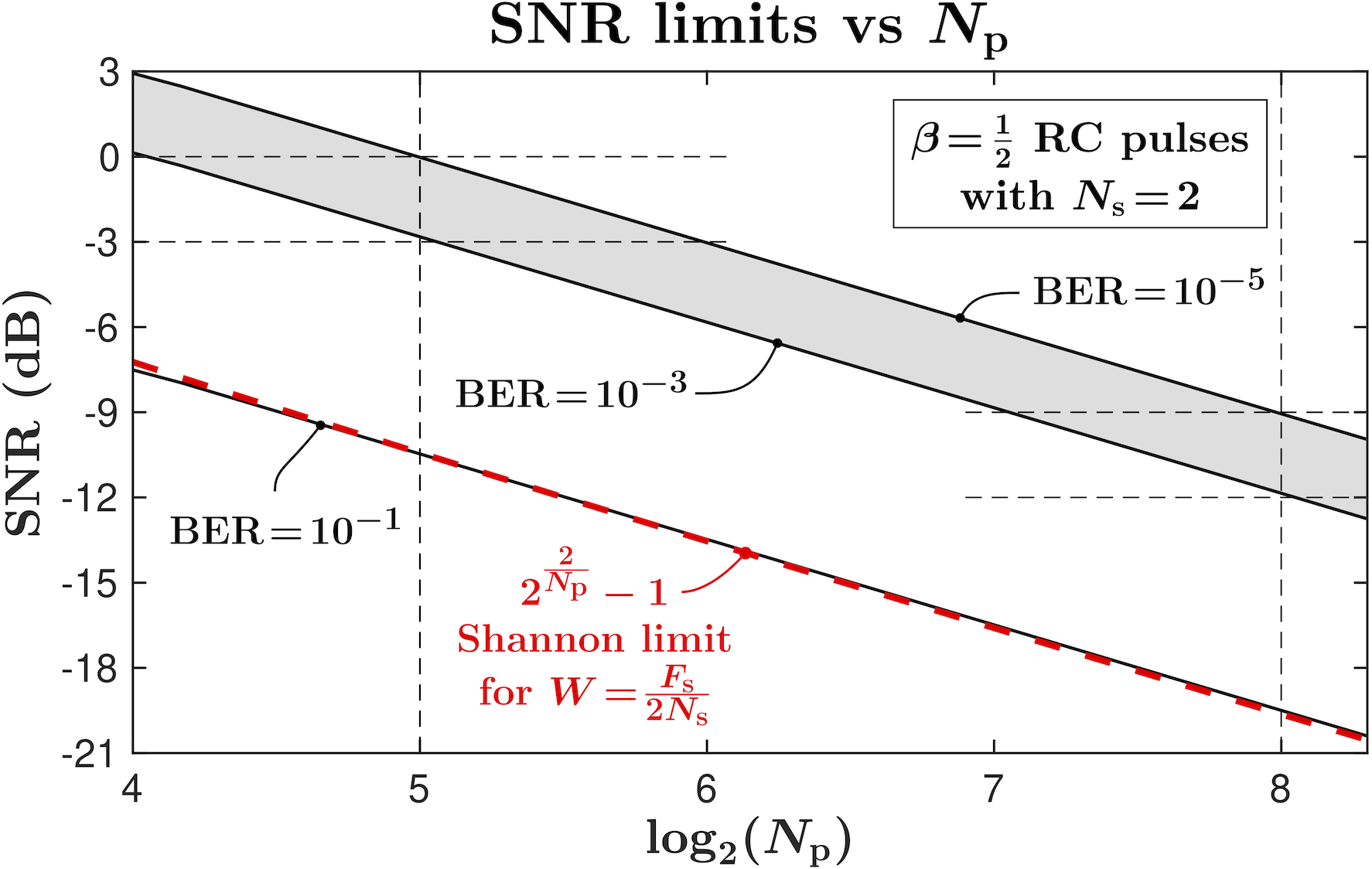}}
\caption{AWGN SNR limits for different BER as functions of samples between pulses for raised-cosine pulses with $\beta\!=\!1/2$ and $N_{\rm s}\!=\!2$.
\label{fig:SNR limits}}
\end{figure}
%--------------------------------------------------------------------------

Fig.~\ref{fig:SNR limits} illustrates the SNR limits for different BER as functions of samples between pulses for RC pulses with $\beta\!=\!1/2$ and $N_{\rm s}\!=\!2$. For example, for the pulses separated by 128~symbol-periods, ${\rm BER}
\!\lesssim\!10{-3}$ is achieved for ${\rm SNR}\!\gtrsim\!-12\,$dB. For comparison, the AWGN Shannon capacity limit~\cite{Shannon49communication} for the bandwidth~$W\!=\!F_{\rm s}/(2N_{\rm s})$, which is the nominal bandwidth of the respective RRC filter, is also shown.

%--------------------------------------------------------------------------
% Asynchronous Detection (Pulse Counting)
\subsection{Asynchronous Detection (Pulse Counting)} \label{subsec:counting}
%--------------------------------------------------------------------------
The asynchronous pulse detection (pulse counting) is discussed in detail in~\cite{Nikitin20steganography}, and it relies on synergistic combination of linear and nonlinear filtering. While the rate limit for pulse counting is approximately an order of magnitude lower than for synchronous pulse detection with a similar BER, pulse counting does not rely on any {\itshape a priori\/} knowledge of pulse arrival times, and can be used as a backbone method for pulse detection. In addition, randomizing the pulse arrival times allows us to more effectively hide the temporal structure of the pulse train, prioritizing security over the data rates. Further, intermittently nonlinear filtering used in combination with synchronous and/or asynchronous pulse detection enables ``layering" of pulse trains with significantly different powers, physical-layer steganography, and ``friendly jamming" applications. However, since synchronous detection enables much higher data rates for the same SNR, the focus of the next section is on the technique that can be used for synchronous detection of pulses in a train with a periodic structure. In practice, both pulse counting and synchronous pulse detection can be used in combination. For example, given a constraint on the total power of the pulse train, counting of relatively rare, higher-magnitude pulses can be used to establish the timing patterns for synchronization, and synchronous detection of smaller, more frequent pulses can be used for a higher data rate.

%--------------------------------------------------------------------------
% FIGURE -- SYNCHRONIZATION
%%% placed here to appear at top of next page
% up to 17.8 cm full width
%--------------------------------------------------------------------------
\begin{figure*}[!t]
\centering{\includegraphics[width=12.4cm]{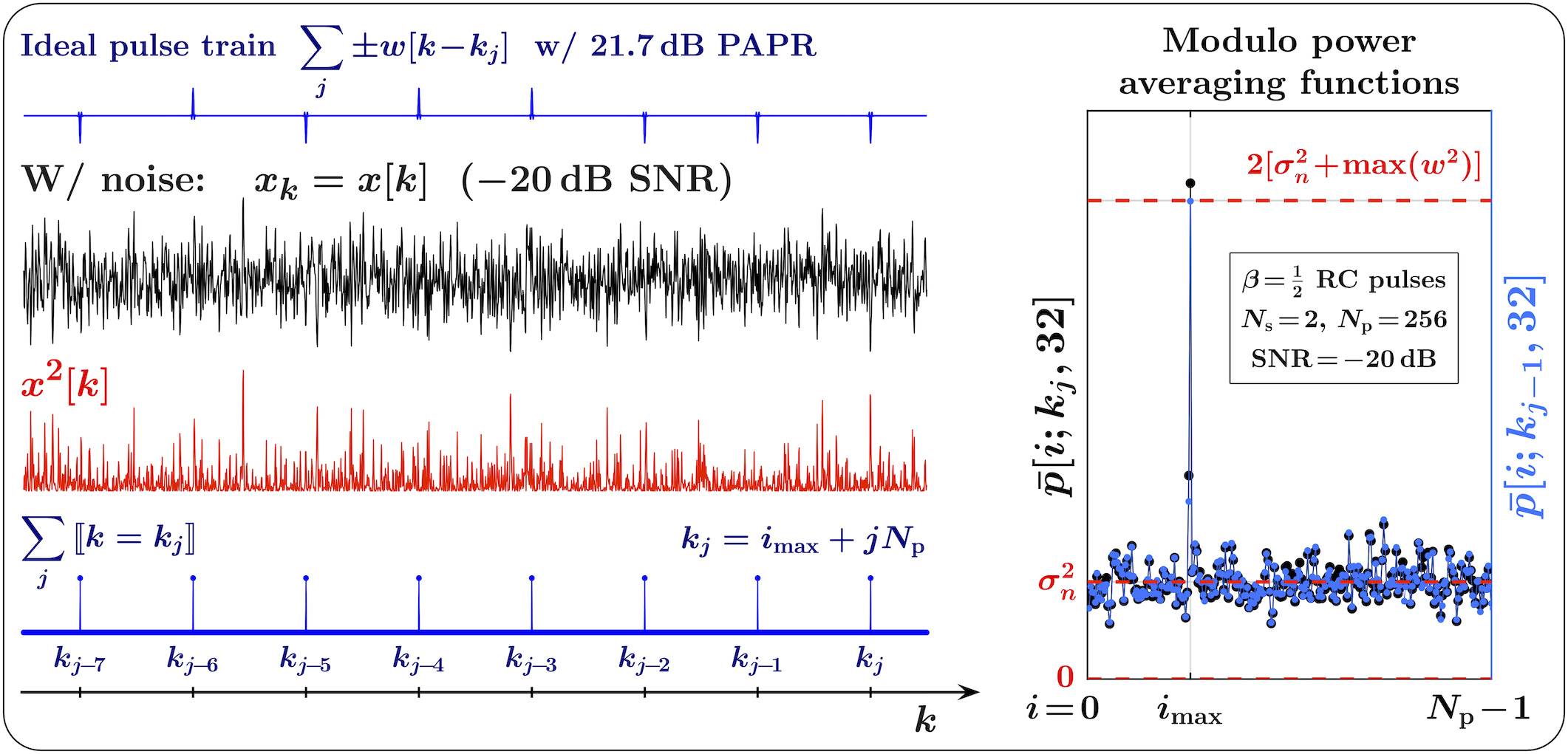}}
\caption{Illustration of synchronization procedure described by (\ref{eq:p bar}) through (\ref{eq:i max}). AWGN ${\rm SNR}=-20\,$dB is chosen to be low, and $M\!=\!32$ respectively high, to emphasize robustness even when ${\rm BER}\approx 1/3$.
\label{fig:synchronization}}
\end{figure*}
%--------------------------------------------------------------------------

%--------------------------------------------------------------------------
% SYNCHRONIZATION
\section{Synchronization} \label{sec:synchronization}
%--------------------------------------------------------------------------
To enable synchronous detection for a train $x[k]$ with the pulses separated by $N_{\rm p}$ samples, the following {\em modulo power averaging\/} (MPA) function can be constructed as an exponentially decaying average of the instantaneous signal power~$x^2[k]$ in a window of size~$N_{\rm p}\!+\!1$:
%-------------------------------------------------------------------
\beginlabel{align}{eq:p bar}
  &\bar{\rm p}[i;k_{j\!-\!1},M] = \frac{M\!-\!1}{M}\, \bar{\rm p}[i;k_{j\!-\!2},M]\\[1ex]
  &+ \frac{1}{M} \sum_k x^2[k] \Ibl k\!\ge\! k_{j\!-\!1}\!-\!N_{\rm p}\Ibr \Ibl k\!\le\! k_{j\!-\!1}\Ibr \Ibl i\!=\!\mod(k,N_{\rm p})\Ibr\,,\nonumber
\end{align}
%-------------------------------------------------------------------
where $k_j$ is the sample index of the $j$-th pulse, and~${M\!>\!1}$. In~(\ref{eq:p bar}), the double square brackets denote the {\em Iverson bracket\/}~\cite{Knuth92two}
%-------------------------------------------------------------------
\beginlabel{equation}{eq:Iverson bracket}
  \Ibl P\Ibr  = \left\{
  \begin{array}{cc}
    \!\! 1 & \mbox{if} \; P \; \mbox{is true}\\
    \!\! 0 & \mbox{otherwise}
  \end{array}\right.,
\end{equation}
%-------------------------------------------------------------------
where $P$ is a statement that can be true or false. Thus the window $k_{j\!-\!1}\!-\!N_{\rm p}\!\le\!k\!\le\! k_{j\!-\!1}$ includes two transmitted pulses, $k_{j\!-\!2}$ and $k_{j\!-\!1}$, and the index~$i$ in $\bar{\rm p}[i;k_{j\!-\!1},M]$ takes the values $i=0,\ldots,N_{\rm p}\!-\!1$. Note that using exponentially decaying average in~(\ref{eq:p bar}) would roughly correspond to averaging $N\!=\!2M\!-\!1$ of such windows. The exponentially decaying average, however, has the advantage of lower computational and memory burden, especially for large~$M$, and faster adaptability to dynamically changing conditions. 

For a sufficiently large~$M$, the peak in $\bar{\rm p}[i;k_{j\!-\!1},M]$ corresponding to the pulses of the pulse train will dominate. Therefore, the index~$k_j$ for sampling of the $j$-th pulse can be obtained as
%-------------------------------------------------------------------
\beginlabel{equation}{eq:kj}
  k_j = i_{\rm max} + j N_{\rm p}\,,
\end{equation}
%-------------------------------------------------------------------
where $i_{\rm max}$ is given implicitly by
%-------------------------------------------------------------------
\beginlabel{equation}{eq:i max}
  \bar{\rm p}[i_{\rm max};k_{j\!-\!1},M] = \max\left( \bar{\rm p}[i;k_{j\!-\!1},M] \right).
\end{equation}
%-------------------------------------------------------------------

Fig.~\ref{fig:synchronization} illustrates this synchronization procedure. The MPA function shown in the right-hand side of the figure is computed according to~(\ref{eq:p bar}). To emphasize the robustness of this synchronization technique even when the bit error rates are very high, the SNR is chosen to be respectively low (${\rm SNR}=-20\,$dB, ${\rm BER}\approx 1/3$).

%--------------------------------------------------------------------------
% FIGURE -- BER vs SNR
%--------------------------------------------------------------------------
\begin{figure}[!t]
\centering{\includegraphics[width=8.6cm]{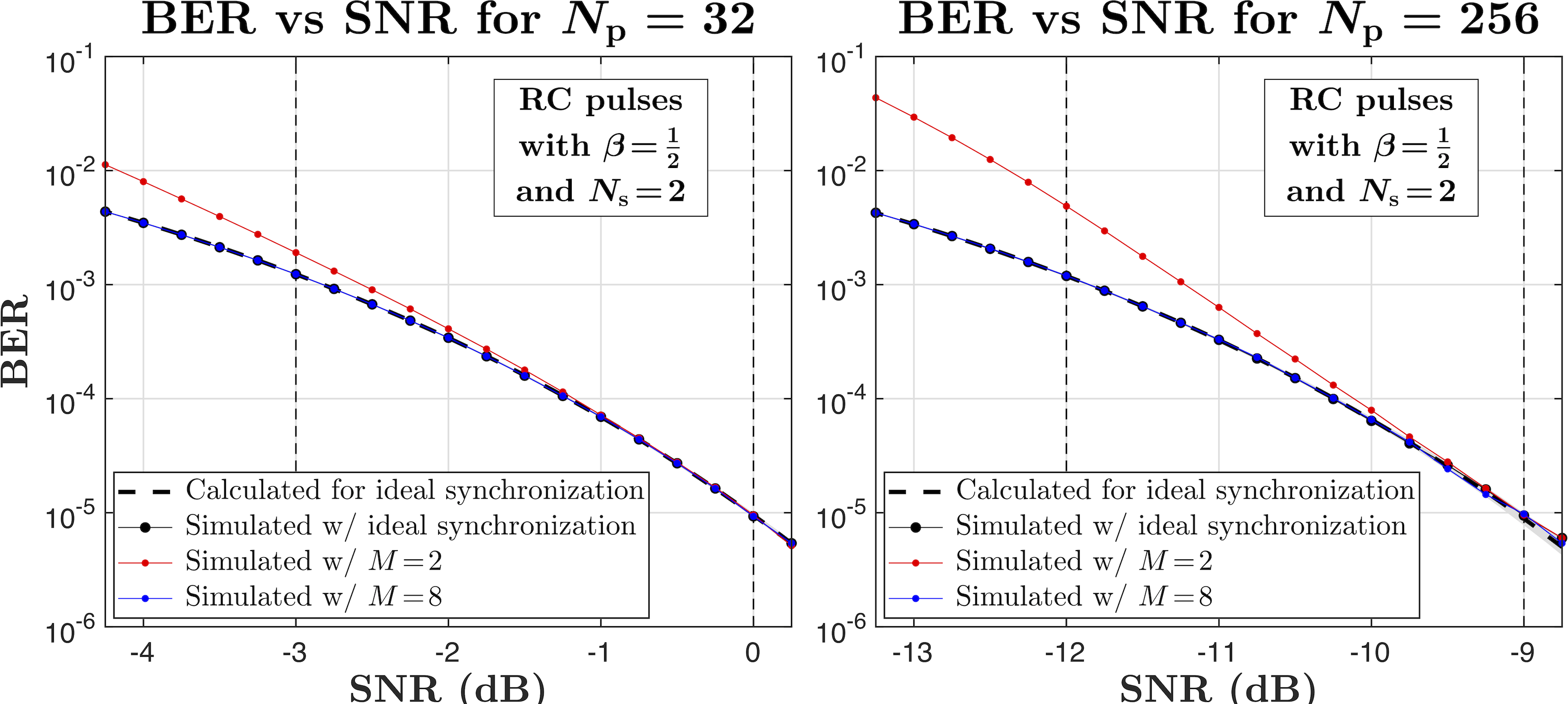}}
\caption{Calculated and simulated BERs as functions of AWGN SNRs for $N_{\rm p}=32$ and $N_{\rm p}=256$. For shown SNR ranges, MPA function with ${M\!=\!8}$ provides reliable synchronization. (Compare with SNR limits in Fig.~\ref{fig:SNR limits}.)
\label{fig:BER vs SNR}}
\end{figure}
%--------------------------------------------------------------------------
%--------------------------------------------------------------------------
% FIGURE -- SYNCHRONIZATION SQ vs ABS
%--------------------------------------------------------------------------
\begin{figure}[!b]
\centering{\includegraphics[width=8.6cm]{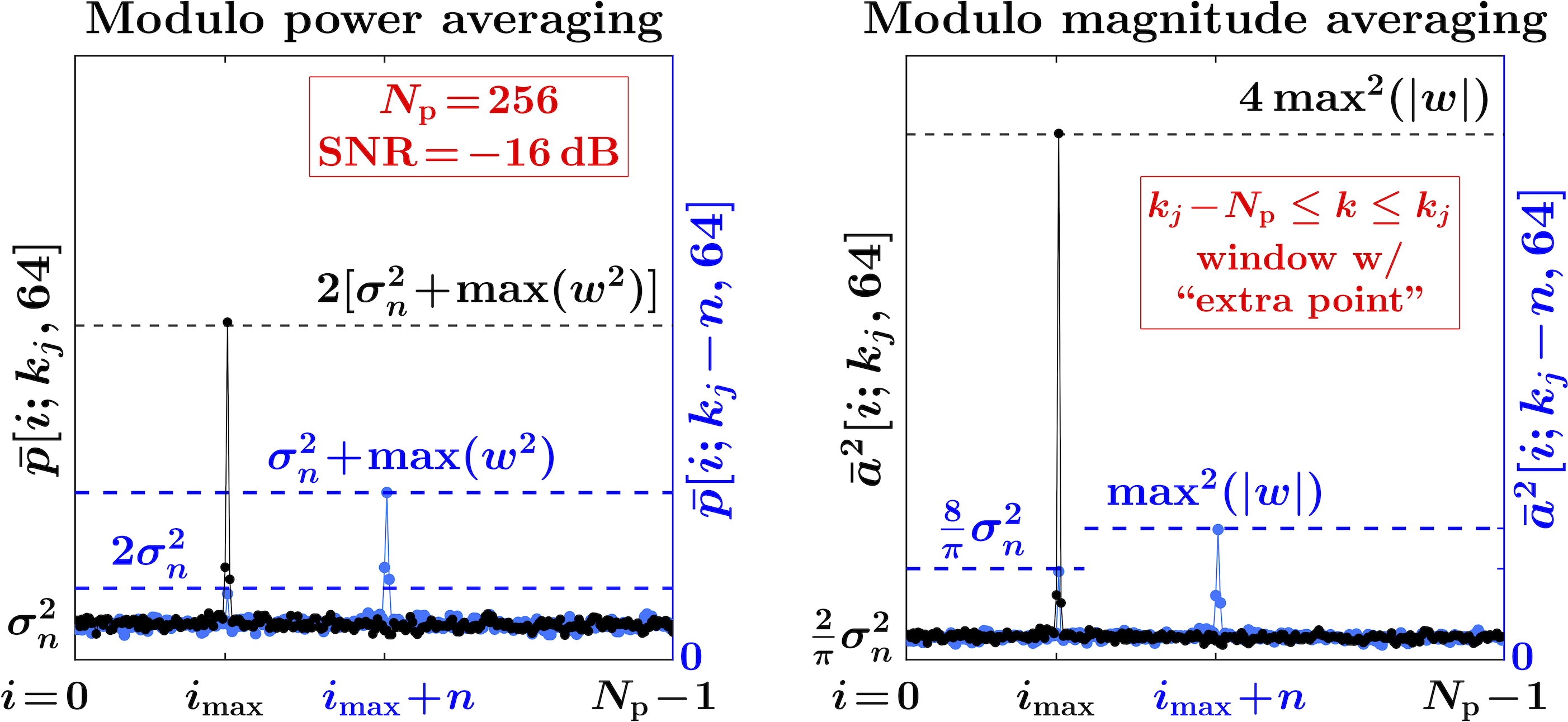}}
\caption{If used in modulo magnitude averaging, ``extra point" significantly increases probability of synchronization failure.
\label{fig:synchronization abs}}
\end{figure}
%--------------------------------------------------------------------------

For the link shown in Fig.~\ref{fig:low SNR}, Fig.~\ref{fig:BER vs SNR} compares the calculated (dashed lines) and the simulated (dots connected by solid lines) BERs, for the ``ideal" synchronization (black dots), and for the synchronization with the MPA function described above. The AWGN noise is added at the receiver input, and the SNR is calculated at the output of the matched filter in the receiver. One can see that for ${M\!=\!2}$ (red dots) the errors in synchronization are relatively high, which increases the overall BER, but the MPA function with ${M\!=\!8}$  (blue dots) provides reliable yet still fast synchronization. The BERs and the respective SNRs in Fig.~\ref{fig:BER vs SNR} are presented for the pulse repetition rates indicated by the vertical dashed lines in Fig.~\ref{fig:SNR limits}.

%--------------------------------------------------------------------------
% Modulo magnitude averaging
\subsection{Modulo magnitude averaging} \label{subsec:MMA}
%--------------------------------------------------------------------------
When a pulse train is used for communications rather than, say, radar applications, reliable synchronization may only need to be achievable for relatively low BER, e.g. ${{\rm BER}\lesssim 1/10}$. Then the following {\em modulo magnitude averaging\/} (MMA) function can replace the MPA function in the synchronization procedure, in order to reduce the computational burden by avoiding squaring operations:
%-------------------------------------------------------------------
\beginlabel{align}{eq:a bar}
  &\bar{\rm a}[i;k_{j\!-\!1},M] = \frac{M\!-\!1}{M}\, \bar{\rm a}[i;k_{j\!-\!2},M]\\[1ex]
  &+ \frac{1}{M} \sum_k |x|[k] \Ibl k\!>\! k_{j\!-\!1}\!-\!N_{\rm p}\Ibr \Ibl k\!\le\! k_{j\!-\!1}\Ibr \Ibl i\!=\!\mod(k,N_{\rm p})\Ibr\,.\nonumber
\end{align}
%-------------------------------------------------------------------
Note that the window $k_{j\!-\!1}\!-\!N_{\rm p}\!<\!k\!\le\! k_{j\!-\!1}$ in~(\ref{eq:a bar}) includes only the $(j\!-\!1)$-th transmitted pulse, instead of two pulses used in~(\ref{eq:p bar}). The reason behind this is illustrated in Fig.~\ref{fig:synchronization abs}, which compares (for AWGN) the MPA function $\bar{\rm p}[i;k_{j\!-\!1},M]$ with the respective squared MMA function $\bar{\rm a}^2[i;k_{j},M]$ computed for the window $k_{j}\!-\!N_{\rm p}\!\le\!k\!\le\! k_{j}$ that includes the ``extra point" (the $(j\!-\!1)$-th pulse). The relatively long averaging ($M\!=\!64$) is used to reduce the variations in the function values due to noise, and to make the comparison with the levels indicated by the dashed lines more apparent.

\begin{figure}[!t]
\centering{\includegraphics[width=8.4cm]{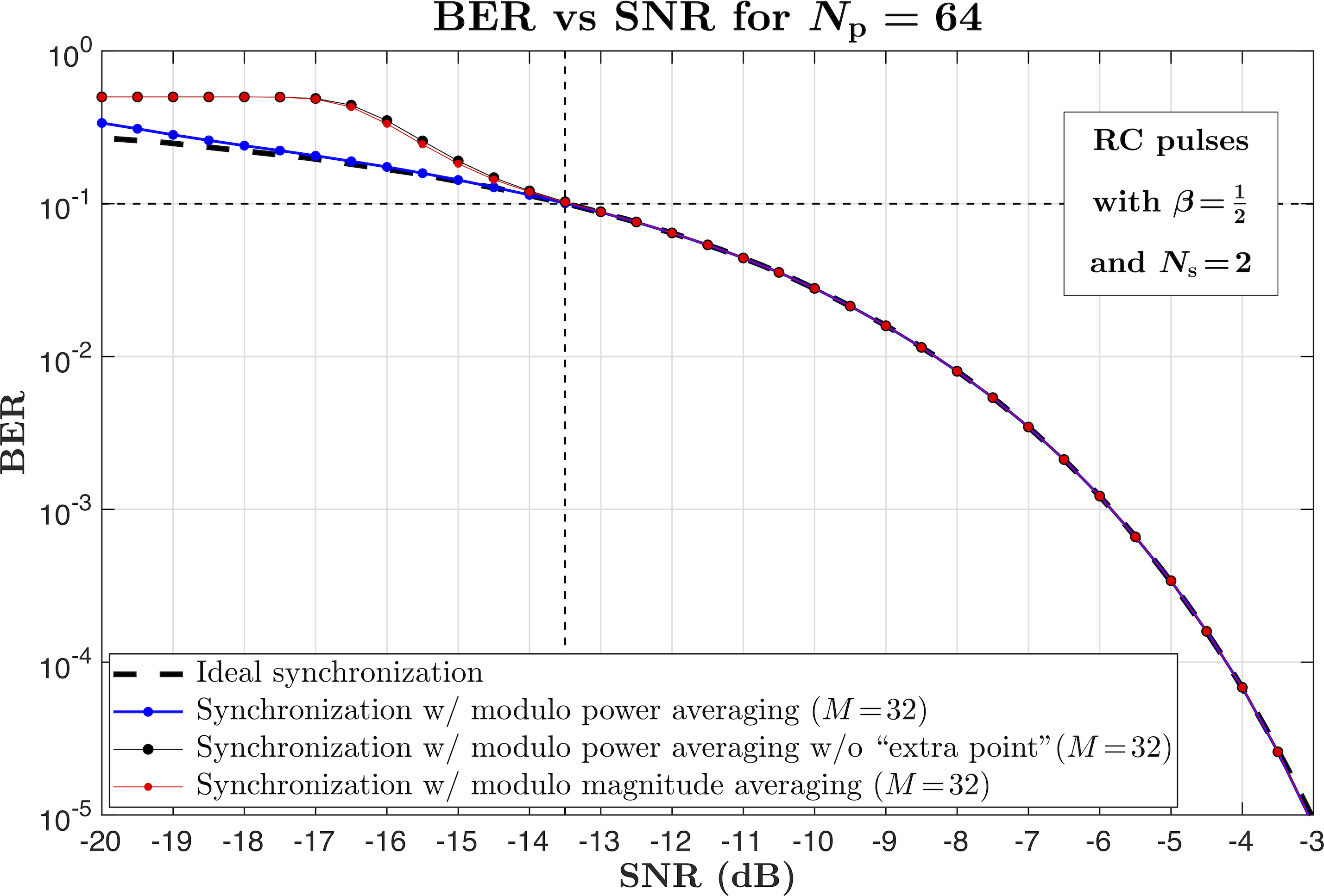}}
\caption{For BER smaller than about $10^{-1}$, less computationally expensive modulo magnitude averaging (e.g. given by~(\ref{eq:a bar})) can be used for synchronization. Modulo power averaging (with ``extra point," e.g. given by~(\ref{eq:p bar})) should be used when reliable synchronization for full BER range is desired.
\label{fig:synchronization comparison}}
\end{figure}

When a correct synchronization has already been obtained, and the maxima are ``locked" at the correct $i_{\rm max}$ values (black dots connected by solid lines), both the MPA and the MMA functions would adequately maintain the position of their maxima. However, an offset in the synchronization (e.g. by $n$ points shown in the figure) significantly more unfavorably affects the margin between the extrema at $i_{\rm max}$ and $i_{\rm max}\!+\!n$ in the MMA function, compared with the MPA function (blue dots connected by solid lines). Thus the ``extra point" may cause the ``failure to synchronize" even at a relatively high SNR, and it should be removed from the calculation of the MMA function. Then, as illustrated in Fig.~\ref{fig:synchronization comparison}, for ${{\rm BER}\lesssim 1/10}$ synchronization with the MMA function $\bar{\rm a}[i;k_{j\!-\!1},M]$ would be effectively equivalent to synchronization with the MPA function $\bar{\rm p}[i;k_{j\!-\!1},M]$. When reliable synchronization for larger BERs is desired (e.g. in timing and ranging applications), then the MPA  given by~(\ref{eq:p bar}) should be used.

%--------------------------------------------------------------------------
% FIGURE -- SYNCHRONIZATION comparison
%--------------------------------------------------------------------------
%--------------------------------------------------------------------------

%--------------------------------------------------------------------------
% CONCLUSION
%--------------------------------------------------------------------------
\section*{Conclusion} \label{sec:conclusion}
\begin{figure}[!t]
\centering{\includegraphics[width=6.6cm]{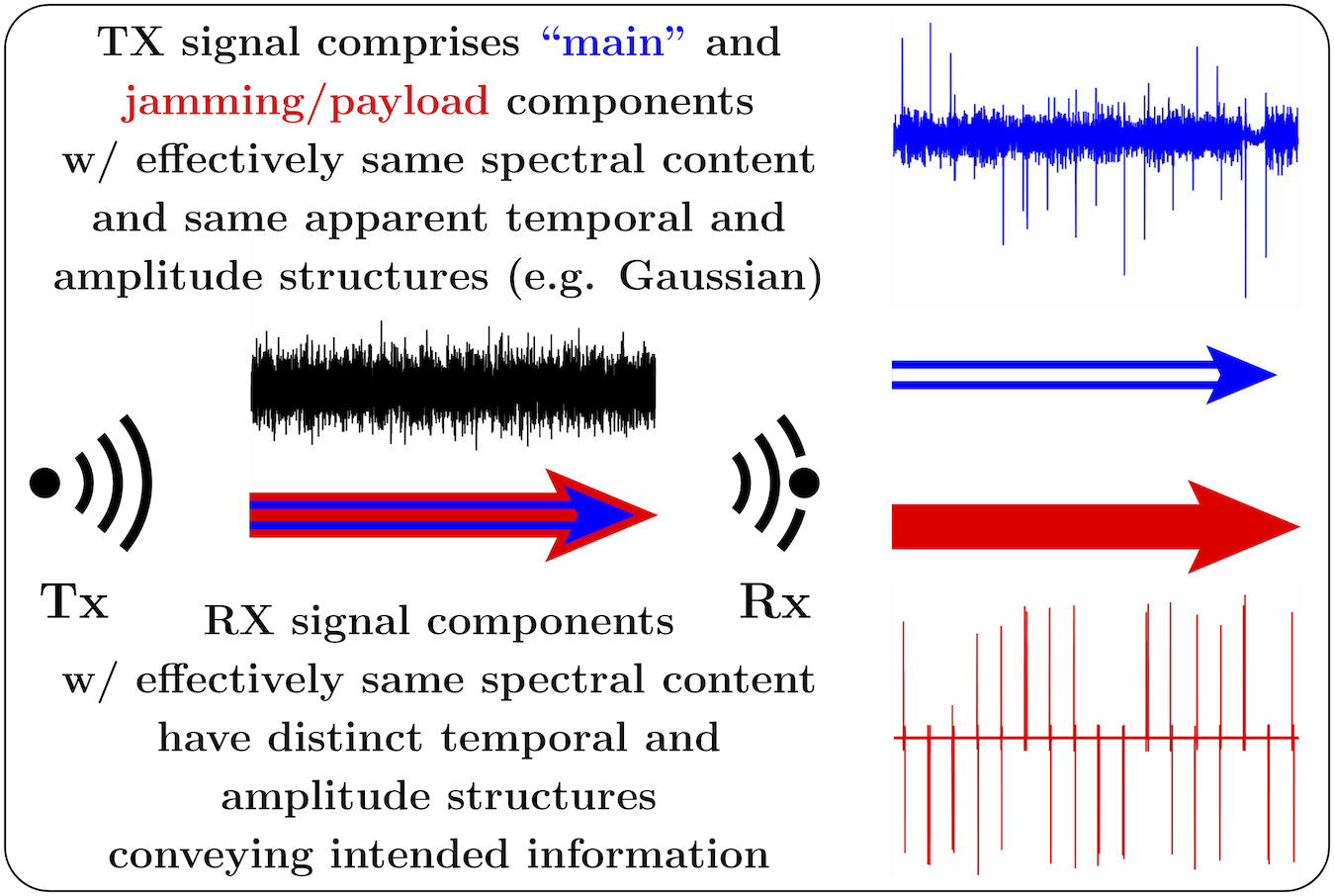}}
\caption{Friendly in-band jamming. 
\label{fig:friendly jamming}}
\end{figure}

%--------------------------------------------------------------------------
Gaining control over the temporal and amplitude structures of the signals in the transmitter and the receiver, especially when combined with nonlinear filtering techniques such as INF~\cite{Nikitin20steganography}, opens up intriguing opportunities in various spectrum sharing and coexistence applications. For example, as schematically illustrated in Fig.~\ref{fig:friendly jamming}, the main message may be transmitted using one of the existing communication protocols (e.g. OFDM), but its temporal and amplitude structure can be obscured by employing a large-TBP filter in the transmitter, e.g., made to be effectively Gaussian. This alone enhances security of the transmission, since the intersymbol interference becomes excessively large and the signal cannot be recovered in the receiver without knowing the pulse shaping filter. Further, a jamming pulse train, disguised as Gaussian by a different large-TBP filter, can be added to the main signal. This jamming signal can have the same spectral content as the main signal, and its power can be sufficiently large (e.g. similar to the main signal) so that the main signal is unrecoverable even if the first pulse shaping filter is known. In the receiver, the jamming pulse train is removed from the mixture by an INF (and decoded, if it itself contains information), enabling the subsequent recovery of the main message without loosing its quality. 

Even a simple single-channel link that is the focus of this paper (see Fig.~\ref{fig:low SNR}) provides appealing practical applications. For example, an existing channel (say, a voice channel in a two-way radio) can be converted into a low-power, lower-rate (say, text) covert channel operating at the same range. This can be accomplished without significant hardware redesign, by modifying only the digital signal processing in both transmitter and the receiver. Alternatively, when transmitted at the original power, such a lower-rate, lower-SNR channel can extend (say, quadruple) the range of the link.

In a broader context, the approach outlined in this paper allows for many practical variations, ranging from simple and easily implementable to more elaborate, highly secure multi-level configurations that would require addressing additional conceptual and implementational challenges.

%--------------------------------------------------------------------------
% FIGURE -- friendly jamming
%--------------------------------------------------------------------------
%--------------------------------------------------------------------------

\section*{Acknowledgment}
%--------------------------------------------------------------------------
The authors would like to thank
James~E. Gilley of BK Technologies, West Melbourne, FL;
Arlie Stonestreet\,\,II of Ultra Electronics ICE, Manhattan, KS,
and Kyle~D. Tidball of Textron Aviation, Wichita, KS,
for their valuable suggestions and critical comments.
This work was supported in part by Pizzi Inc., Denton, TX 76205 USA.
%--------------------------------------------------------------------------

%--------------------------------------------------------------------------
% REFERENCES
%--------------------------------------------------------------------------

%--------------------------------------------------------------------------
\end{document}